\documentclass[12pt]{article}
\begin{document}
\title{Universal criterion for black hole stability}
\author{Ashok Chatterjee\footnote{email: ashok.chatterjee@saha.ac.in} 
and Parthasarathi
Majumdar\footnote{email: parthasarathi.majumdar@saha.ac.in} \\Theory
Group,  Saha Institute of Nuclear Physics, Kolkata 700 064, India.}
\maketitle
\begin{abstract}
It is shown that a non-rotating macroscopic black hole with very large horizon 
area can remain in stable thermal equilibrium with Hawking radiation provided
{\it its mass, as a function of horizon area, exceeds its microcanonical entropy,
i.e., its entropy when isolated, without thermal radiation or accretion,
and having a constant horizon area} (in appropriate units). The
analysis does not use properties of specific classical spacetimes, but
depends only on the plausible assumption that the mass is a function
of the horizon area for large areas.
\end{abstract}

Black holes whose spacetimes are flat at infinity, like the
Schwarzschild spacetime, exhibit a thermal 
instability: they either radiate or accrete thermally without
limit. In the former case they may evaporate away completely
\cite{haw1,haw2} with
possible loss of information. In the latter case they may turn into 
supermassive black holes the likes of which have recently been claimed
to have been observed. The thermal instability of asymptotically flat
black holes has been
variously attributed to a negative heat capacity or a superexponential
density of states leading to a diverging canonical partition
function. In contrast, black hole spacetimes which asymptotically
have constant negative curvature (anti-de Sitter or AdS ) have special
properties enabling them to coexist in stable thermal equilibrium
with a bath of their own radiation \cite{hawp} . Are there other black holes
which may be in stable equilibrium with radiation ? It is clear that
semiclassical approaches employing explicitly classical
background metrics will have limited applicability in unravelling issues like
this, since black hole thermodynamics has origins most definitely
grounded in the quantum nature of spacetime \cite{bek}. 

A background-independent canonical quantization of general relativity, 
known as Loop Quantum Gravity (LQG) \cite{al1} has yielded a fairly complete 
understanding of the entropy of {\it isolated} black
holes \cite{abck, km, dkm} which are neither radiating nor accreting thermally,
and are therefore of fixed horizon area ${\cal A}$. For
non-rotating isolated black holes with large ${\cal A} >>
l_{Planck}^2$ ($l_{Planck}$ is the Planck length $(G
\hbar/c^3)^{1/2}$) the microcanonical entropy $S_{MC}({\cal A})$ is
given by an inifinite series in inverse powers of horizon area
\cite{km,dkm},
\begin{eqnarray}
S_{MC}({\cal A})~=~S_{BH}~-{\xi \over 2}~\log S_{BH} ~+~ const.~+~
O(S_{BH}^{-1}) ~, ~\label{smc}
\end{eqnarray}
where, $S_{BH} \equiv {\cal A}/4l_{Planck}^2 $ is the
Bekenstein-Hawking entropy, $\xi = 3$, if the residual
gauge invariance on a spatial slice of the horizon (leftover from
local Lorentz invariance) is $SU(2)$, and $\xi=1$ if the gauge group
is $U(1)$. The subleading terms in eq. (\ref{smc}) are all finite and
unambiguously calculable. The only ambiguity involves the
Barbero-Immirzi parameter $\gamma$ which is fixed by fitting the
leading area term to $S_{BH}$. The log(area) corrections are clearly
independent of this ambiguity. In obtaining this result, crucial use
has been made of the fact that in LQG, geometrical quantities like
area and volume are represented by self-adjoint operators acting on
the (kinematical) Hilbert space, which can be shown to have a discrete
spectrum \cite{al2}. For very large areas (in units of Planck area), the area
spectrum can be shown to be characterized by an integer, ${\cal A}_n
\sim l_{Planck}^2~n$ with $n \gg 1$. 

The microcanonical approach is however inadequate from a physical
perspective since a black hole does necessarily radiate and accrete
thermally. To handle this dynamical situation, Loop Quantum Gravity
per se is not very useful, because of long-standing difficulties
inolving the quantum Hamiltonian constraint \cite{al1}. In absence of a
direct quantization of a dynamical horizon, the present authors have
adopted an indirect heuristic approach based on the standard canonical 
ensemble of equilibrium statistical mechanics. This has been applied to 
nonrotating black holes assumed to be in contact with their radiation
bath \cite{cm1,cm2,cm3}. The horizon is assumed to be an inner boundary of
spacetime. With this assumption, and the fact that quantum states
corresponding to bulk three dimensional space (on a spatial slice) are
annihilated by the quantum Hamiltonian operator, the partition
function essentially reduces to computing the state sum over the
horizon states. This computation is performed in the saddle point
approximation \cite{cm1,cm2}, including the Gaussian fluctuations around
the saddle point (identified here with the classical mass $M$), and
one obtains
\begin{eqnarray}
Z_{hor}~\simeq~ \exp \left\{ S_{MC}(M) - \beta M - \log|{dE \over 
dx}|_{E=M} \right \}~\left [{\pi \over -S_{MC}''(M)} \right]^{1/2}  
~. \label{sad}
\end{eqnarray}
In this equation, $x$ is the continuum variable which has replaced the
area quantum number $n$ introduced earlier for large areas. The
quantity $dE/dx|_M = M'({\cal A}) \cdot  const.$ since $d{\cal A}/dx =
const.$ from the area spectrum for large areas (with prime indicating
a derivative with respect to the argument). Notice that we have made
the tacit assumption that the black hole mass is a function of the
area. This is not really an assumption for many classical general 
relativistic black holes for asymptotically large areas. Such a functional 
dependence is plausible even in LQG given that the bulk
Hamiltonian can be related to the volume operator \cite{thie} in LQG. 
We also note that the quantity in square brackets in eq. (\ref{sad}) is
the contribution of Gaussian fluctuation around the saddle point
at $E=M$. 

It is interesting that this now leads to the following
canonical entropy for non-rotating black holes \cite{cm2}
\begin{eqnarray}
S_{can}~=~S_{MC}({\cal A})~-~\frac12 \log \Delta~, \label{cane} 
\end{eqnarray}
where
\begin{eqnarray}
\Delta~\equiv~[{\cal A}'(x)]^2~\left[ S_{MC}'({\cal A}) ~{M''({\cal A})
\over M'({\cal A})}~-~S_{MC}''({\cal A}) \right]~.
\label{delta}
\end{eqnarray}
Thus, the canonical entropy is expressed in terms of the
microcanonical entropy for an average large horizon area, and the
mass which is also a function of the area. Clearly, stable 
equilibrium ensues so long as $\Delta > 0$. 

Additional support for this condition can be gleaned by considering
the thermal capacity of the system, using the standard relation 
\begin{eqnarray}
C({\cal A}) \equiv {dM \over dT} = {M'({\cal A}) \over T'({\cal A})},
\label{hea}
\end{eqnarray}
with $T$ being derived from the microcanonical entropy $S_{MC}({\cal
A})$, and hence a function of ${\cal A}$. One obtains for the heat
capacity the relation
\begin{eqnarray}
C({\cal A})~=~\left[{M'({\cal A}) \over T({\cal A}) {\cal A}'(x)}
\right]^2~\Delta^{-1} ~, \label{hc}
\end{eqnarray}
so that $C > 0$ if only if $\Delta > 0$. Since the positivity of the
heat capacity is certainly a necessary condition for stable thermal
equilibrium, it is gratifying that an identical criterion emerges
for $\Delta$ as found from the canonical entropy (\ref{cane}). 

Using now eq, (\ref{delta}) for the expression for $\Delta$, the 
criterion
for thermal stability of non-rotating macroscopic black holes is then
easily seen to be
\begin{eqnarray}
M({\cal A}) ~>~ S_{MC}({\cal A}) ~
\label{crit}
\end{eqnarray}
as already mentioned in the summary. We have been using units in which
$G=\hbar=c=k_B=1$. If we revert back to units where these constants
are not set to unity, the lower bound eq. (\ref{crit}) can be re-expressed
as 
\begin{eqnarray}
M({\cal A}) ~>~ \left({\hbar c \over G k_B^2} \right)^{1/2}
S_{MC}({\cal A}) ~. \label{crit2}
\end{eqnarray}
We remind the reader that in
contrast to semiclassical approaches based on specific properties of
classical metrics, our approach incorporates crucially the
microcanonical entropy generated by quantum spacetime fluctuations
that leave the horizon area constant. Apart from the plausible
assumption of the black hole mass being dependent only on the horizon
area, no other assumption has been made to arrive at the result. Even
so, it subsumes most results based on the semiclassical approach. It
also supercedes our earlier assay \cite{cm2} based on an assumption of a
power law functional dependence of the mass on the area.  

As a byproduct of the above analysis, the canonical entropy for stable
black holes can be expressed in terms of the Bekenstein-Hawking
entropy $S_{BH}$ as
\begin{eqnarray}
S_{can}~&=&~S_{BH} -\frac12 (\xi- 1) \log S_{BH} \nonumber \\
~&-&~\frac12 \log
\left[ {S'_{MC}({\cal A}) ~ M''({\cal A}) \over S_{MC}''({\cal A})
 M'({\cal A})} \right] ~.\label{sc}
\end{eqnarray}  
For any smooth $M({\cal A})$, one can truncate its power series
expansion in ${\cal A}$ at some large order and show that the quantity
in square brackets in eq. (\ref{sc}) does not contribute to the $\log
(area)$ term, so that
\begin{eqnarray}
S_{can}~=~S_{BH}~-\frac12~(\xi-1)~\log S_{BH}~+~const.~+~O(S_{BH}^{-1}) ~. 
\label{scan}
\end{eqnarray}
The interplay between constant area quantum spacetime fluctuations
and thermal fluctuations is obvious in the coefficient of the
$log(area)$ term where the contribution due to each appears with a
specific sign. It is not surprising that the thermal fluctuation
contribution increases the canonical entropy. The cancellation
occurring for horizons on which a residual $U(1)$ subgroup of $SU(2)$
survives, because of additional gauge fixing by the boundary conditions
describing an isolated horizon \cite{al1}, may indicate a possible
non-renormalization theorem, although no special symmetry like
supersymmetry has been employed anywhere above. It is thus generic for
all non-rotating black holes, including those with electric or dilatonic
charge. One would expect the result to hold also for rotating black holes, 
as well, although the details of the microcanonical entropy for such
black holes have not yet been worked out. 

While this letter restricts attention to thermal fluctuations of area
due to energy fluctuations alone, the stability criterion (\ref{crit})
holds when in addition thermal fluctuations of electric charge are
incorporated within a grand canonical ensemble. The result for the
grand canonical entropy is however somewhat different from that given
above (\ref{scan}) when charge fluctuations are included \cite{cm4}.


\begin{thebibliography}{99}
\bibitem {haw1} Hawking S.W., Black hole explosions ? {\it Nature} {\bf 248}, 30-31 
(1974). 
\bibitem {haw2} Hawking S.W., Particle creation by black holes, {\it Commu. Math. Phys.},
{\bf 43}, 199-222 (1975).
\bibitem {hawp} Hawking, S. W. and Page, D. N., Thermodynamics of black holes in
anti-de Sitter space, {\it Commu. Math. Phys.} {\bf 87,} 577 (1983).
\bibitem {bek} Bekenstein, J. D., Black holes and entropy, {\it Phys. Rev.} {\bf
D7}, 2333-2346 (1973).
\bibitem {al1} Ashtekar, A. and J. Lewandowski, Background Independent Quantum
Gravity : a Status Report, {\it Class. Quant. Grav.} {\bf 21}, R53
(2004).   
\bibitem {abck} Ashtekar A., Baez, J. C. and Krasnov, K., Quantum Geometry of
Isolated Horizons and Black Hole Entropy, {\it Adv. Theor. Math. Phys.}, 
{\bf 4}, 1-94 (2000).
\bibitem {km} Kaul, R. K. and Majumdar, P., Logarithmic correction to the
Bekenstein-Hawking entropy, {\it Phys. Rev. Lett.} {\bf 84}, 5255-5257
(2000).
\bibitem {dkm} Das, S., Kaul, R. K. and Majumdar, P., A new holographic entropy
bound from quantum geometry, {\it Phys. Rev.} {\bf D63} 044019
(2001). 
\bibitem {al2} Ashtekar, A. and Lewandowski, J., Quantum theory of geometry I :
Area operators, {\it Class. Quant. Grav.} {\bf 14}, A55-A81 (1997).
\bibitem {cm1} Chatterjee, A. and Majumdar, P., Black hole entropy, quantum
versus thermal fluctuations, {\it ArXiv:gr-qc/0303030}. 
\bibitem {cm2} Chatterjee, A. and Majumdar, P., Universal canonical black hole
entropy, {\it Phys. Rev. Lett.} {\bf 92}, 141301 (2004). 
\bibitem {cm3} Chatterjee, A. and Majumdar, P., Universal canonical entropy for
gravitating systems, {\it Pramana}, {\bf 63}, 851-858 (2004). 
\bibitem {thie} Thiemann, T., Anomaly-free formulation of non-perturbative four
dimensional Lorentzian quantum gravity, {\it Phys. Lett.} {\bf B380}
257-264 (1996). 
\bibitem {cm4} Chatterjee, A. and Majumdar, P., Mass and charge fluctuations
and black hole entropy, {\it Phys. Rev.} {\bf D 71}, 024003 (2005). 
\end{thebibliography}
\end{document}